\documentclass[10pt,letterpaper]{article}
\usepackage{opex3}
\usepackage{amsmath}
\usepackage{amssymb}

\begin{document}

\title{Modified field enhancement in plasmonic nanowire dimers due to nonlocal response}

\author{Giuseppe Toscano,$^{1}$ S{\o}ren Raza,$^{1,2}$ Antti-Pekka Jauho,$^{3}$\\   N. Asger Mortensen,$^{1}$ and Martijn Wubs$^{1}$}

\address{$^{1}$DTU Fotonik, Department of Photonics Engineering, Technical University of Denmark,
DK-2800 Kongens Lyngby, Denmark}
\address{$^{2}$DTU Cen, Center for Electron Nanoscopy, Technical University of Denmark, DK-2800
Kongens Lyngby, Denmark}
\address{$^{3}$DTU Nanotech, Department of Micro- and Nanotechnology, Technical University of
Denmark, DK-2800 Kongens Lyngby, Denmark}
\email{asger@mailaps.org, mwubs@fotonik.dtu.dk} 

\begin{abstract}
We study the effect of nonlocal optical response on the optical properties of metallic nanowires, by numerically implementing the hydrodynamical Drude model for arbitrary nanowire geometries. We first demonstrate the  accuracy of our frequency-domain finite-element implementation by benchmarking it in a wide frequency range against analytical results for the extinction cross section of a cylindrical plasmonic nanowire. Our main results concern more complex geometries, namely cylindrical and bow-tie nanowire dimers that  can strongly enhance optical fields.
For both types of dimers we find
that nonlocal response can strongly affect both the field enhancement in between the dimers and their respective extinction cross sections. In particular, we give examples of maximal field enhancements near hybridized plasmonic dimer resonances that are still large but nearly two times smaller than in the usual local-response description. At the same time, for a fixed frequency the field enhancement and cross section can also be significantly more enhanced in the  nonlocal-response model.
\end{abstract}



\section{Introduction}
In plasmonics~\cite{Boardman:1982a,Maier:2007a,Gramotnev:2010a}, subwavelength metal structures are used to confine and enhance~\cite{Aizpura:2005a,Xiao:2008a}, guide~\cite{Wei:2010a,Chen:2011a}, and scatter~\cite{Rockstuhl:2004a,Muskens:2007a,Giannini:2007a,Shegai:2011a} light. The goal to measure and control light at ever smaller length scales drives the research field towards true nanoplasmonics. It is then a natural question to ask: down to which sizes is the optical description of the metal solely in terms of its bulk dielectric function still accurate?

One phenomenon beyond this usual description that becomes important is nonlocal optical response~\cite{Boardman:1982a,Abajo:2008a,McMahon:2009a,Raza:2011a,David:2011a}: the fact that not only light but also moving electrons in the metal transport energy. Here we will focus on effects due to nonlocal response, using the linearized hydrodynamic Drude model~\cite{Boardman:1982a,Ruppin:2001a,VilloPerez:2009a,Raza:2011a}. Compared to the usual local-response Drude theory for free electrons, this hydrodynamic theory has the Fermi velocity of the electrons as an additional parameter.

 Nowadays, simulating local-response plasmonic properties has almost become a standard task, even for complex geometries, thanks to the availability of advanced numerical methods and dedicated software. Because of the continuing success in fabricating ever smaller plasmonic nanostructures, the nonlocal response will become increasingly important. This has stimulated us to develop a similar reliable and easy-to-use numerical tool also for nonlocal response, and apply it to geometries where we expect effects of nonlocal response to be significant, as presented here. The nonlocal calculations are numerically more challenging, since the Fermi wavelength of the order of $k_{\rm F}^{-1}\sim 1\,{\rm nm}$ enters as a new length scale of longitudinal waves~\cite{Boardman:1982a,Ruppin:2001a,VilloPerez:2009a,Raza:2011a}. The computational grid with typical separations $\Delta x$ should resolve not just the sub-wavelength features, but rather sub-Fermi-wavelength features of the geometry of typical size $L$ and field distributions for optical wavelengths $\lambda$. The numerical grid size $\Delta x$ must be smaller than all physical length scales in our study, and the latter satisfy $k_{\rm F}^{-1}<L\ll\lambda$.

 The core of this article is a numerical study of nonlocal-response effects when light scatters off nanoplasmonic dimer structures, which are archetypical structures to study both field enhancement~\cite{Aizpura:2005a,Aubry:2010a}, scattering~\cite{Rockstuhl:2004a,Muskens:2007a},  and hybridization of plasmonic resonances~\cite{Kottmann:2001a,Prodan:2003a,Brandl:2005a,Davis:2010a}. Important is also that  dimers can display resonances in the visible~\cite{Koh:2011a} even when their two constituents, taken separately, would not.
As is known from local-response hybridization theory, hybridization energies grow as dimer distances are reduced. Here we study how nonlocal response affects his behavior. Our study of nanowire dimers complements recent work on dimers of nanospheres~\cite{Abajo:2008a,Yannopapas:2008a,David:2011a}.  Here we present results for dimer separations only down to $1\,{\rm nm}$, because for smaller separations, quantum effects not taken account into our model are predicted to strongly reduce hybridization energies~\cite{Zuloaga:2009a}. We focus solely on extinction cross sections and field enhancements that can be probed with light, leaving for later study the dark modes of the nanowire dimers that could be seen in electron energy loss spectroscopy~\cite{Koh:2011a,Nicoletti:2011a} or cathodoluminescence experiments~\cite{Barnard:2011a}.

The structure of this article is as follows. In Sec.~\ref{Sec:theory} we introduce the theoretical formalism, and the numerical implementation in Sec.~\ref{Sec:Implementation}. In Sec.~\ref{Sec:1cylinder} we benchmark our implementation of the hydrodynamical Drude model against the analytically solvable problem of the scattering off a single cylindrical nanowire, and also study the size dependence of nonlocal effects. In Sec.~\ref{Sec:2cylinders} we compare nonlocal response against local response for dimers consisting of two such cylindrical nanowires where field enhancement occurs in the open cavity  between the cylinders. Then   in Sec.~\ref{Sec:bowtie} we do the same for bow-tie dimer nanowires, where field enhancement occurs near the almost touching sharp tips of the triangles. Our conclusions are given in Sec.~\ref{Sec:Conclusions}, and details on our numerical calculations in the Appendix.

\section{Theoretical formalism}\label{Sec:theory}
The usual local-response dielectric function of realistic metals is the sum of a Drude free-electron response  plus interband effects~\cite{Maier:2007a,Shegai:2011a}. In the hydrodynamic description, only the free-electron response which is modified while the other (interband) effects are unaltered. In the Maxwell wave equation for the electric field $\mathbf E$, the free-electron response is described by a current density $\mathbf J$, while the rest of the optical response is modeled with a local, usually spatially piecewise constant,  dielectric function $\varepsilon_{\rm other}({\bf r},\omega)$. We are interested in the linear optical response, and the linearized hydrodynamical model then leads to coupled equations for the electrical field $\mathbf E$ and the current density $\mathbf J$~\cite{Boardman:1982a,Raza:2011a}
\begin{subequations}
\label{eq:coupledequations}
\begin{equation}
{\mathbf\nabla}\times{\mathbf\nabla}\times{\mathbf E}({\bf r},\omega)=\varepsilon_{\rm other}({\bf r},\omega)\frac{\omega^2}{c^2}{\mathbf E}({\bf r},\omega) +
i\omega\mu_0 {\mathbf J}({\bf r},\omega),\label{eq:Maxwell}
\end{equation}
\begin{equation}
	\frac{\beta^2}{\omega\left(\omega+i/\tau_{\rm Drude}\right)} {\mathbf \nabla} \left[ {\mathbf \nabla} \cdot {\mathbf J}({\bf r},\omega) \right] +
 {\mathbf J}({\bf r},\omega) = \sigma({\bf r},\omega) {\mathbf E}({\bf r},\omega).
\label{eq:lmotion}
\end{equation}
\end{subequations}
 Also in linear response, the static electron density  is homogeneous inside the metal while it is vanishing outside. In Eq.~(\ref{eq:lmotion}), this electron density has been parameterized by the spatially piecewise constant AC conductivity $\sigma({\bf r},\omega)$ and $\tau_{\rm Drude}$ is the Drude damping time that also occurs in the Drude local-response dielectric function $\varepsilon(\omega)=1-\omega_{p}^2/[\omega(\omega+i/\tau_{\rm Drude})]$, where $\omega_{\rm p}$ is the plasma frequency. We take parameters for gold, namely the plasma frequency $\hbar \omega_{\rm p} = 8.812\,{\rm eV}$, Drude damping  $\hbar/\tau_{\rm Drude} = 0.0752\,{\rm eV}$, and Fermi velocity $v_{\rm F} = 1.39 \times 10^{6}\,{\rm m/s}$, the same values as in Refs.~\cite{McMahon:2009a,Raza:2011a}.
In the Thomas-Fermi model and for $\omega_{\rm p}\tau_{\rm Drude} \gg 1$, the nonlocal-response parameter $\beta$ is proportional to the Fermi velocity through $\beta = \sqrt{3/(D+2)} v_{\rm F}$. Here $D$ is the number of spatial dimensions from the point of view of the electron dynamics, which is the number of dimensions that are not quantum confined~\cite{Fetter:1973a}.
For $\beta\rightarrow 0$ we recover the local-response model where the dynamics is governed by Ohm's law with ${\mathbf J}=\sigma{\mathbf E}$.
We leave out the additional complexity of interband effects by taking $\varepsilon_{\rm other}({\bf r},\omega)\equiv 1$ in Eq.~(\ref{eq:Maxwell}). The  interband effects could be taken into account as well, following Refs.~\cite{Abajo:2008a,McMahon:2009a,David:2011a}.

As stated in the Introduction, we do not consider the `spill-out' of the electron density at the metal surface leading to quantum tunneling, as described by microscopic  many-body calculations~\cite{Zuloaga:2009a,Oeztuerk:2011a}. As an immediate consequence of this approximation, the normal component of the current ${\bf J}$ vanishes at the surface of the metal volume(s). We proceed to solve the coupled equations~(\ref{eq:coupledequations}) self-consistently, with the usual Maxwell boundary conditons plus the additional boundary condition of the vanishing normal component of the current ${\bf J}$ at the metal surface(s).  For further details of the model and the appropriate boundary conditions we refer to our recent theoretical work~\cite{Raza:2011a}.

\section{Computational method and implementation}\label{Sec:Implementation}
We first discuss the light extinction properties of infinitely long nanowire structures surrounded by free space. Since  the electrons are confined in two directions but not quantum-confined in the nanowires, we should take $D=3$ whereby the nonlocal parameter  becomes $\beta=\sqrt{3/5}v_{\rm F}$~\cite{Fetter:1973a}. As the light source, we take a monochromatic plane wave incident normal to the wire axis and with an in-plane polarized electrical field (TM polarization).

We solve Eq.~(\ref{eq:coupledequations}) numerically with the aid of a finite-element method (FEM).
To obtain a reliable and flexible implementation,  we have built it as a nonlocal-response extension of a commercially available multi-purpose code for local-response (the RF module of Comsol Multiphysics, version 4.1). The FEM is known to have a high ability to handle complex geometries, and to accurately model small surface details, such as gaps and
tips~\cite{Gockenbach:2006a}. Furthermore, retardation is automatically taken fully into account, which is important according to the very recent study by David and Garc{\'i}a de Abajo~\cite{David:2011a}.

The translational invariance of the structure in one spatial dimension simplifies the calculation, since the  calculation domain (or `grid') becomes two-dimensional. Within this plane, the wire geometry can be chosen at will. We imbed the 2D-projection of our metallic nanostructure into a
square computational domain.  Perfectly-matched layers (PML) at the edges of this domain mimic the reflectionless coupling to the surrounding free space.  For the meshing of the geometry we take advantage of the built-in algorithm of the software, paying special attention to mesh refinement needed to account for surface effects and abrupt changes in the surface topography.

Our code runs on a pc. For details about our hardware implementation we refer to the Appendix. The code was first successfully tested in the limit $\beta\rightarrow 0$ [see Eq.~(\ref{eq:lmotion})], where it correctly reproduces the local scattering response of various standard problems. Below, we report our benchmarking against analytical results, both for local and for nonlocal response.

\section{Benchmark problem: a single cylindrical nanowire}\label{Sec:1cylinder}
\begin{figure}[t]
\centering 
\includegraphics{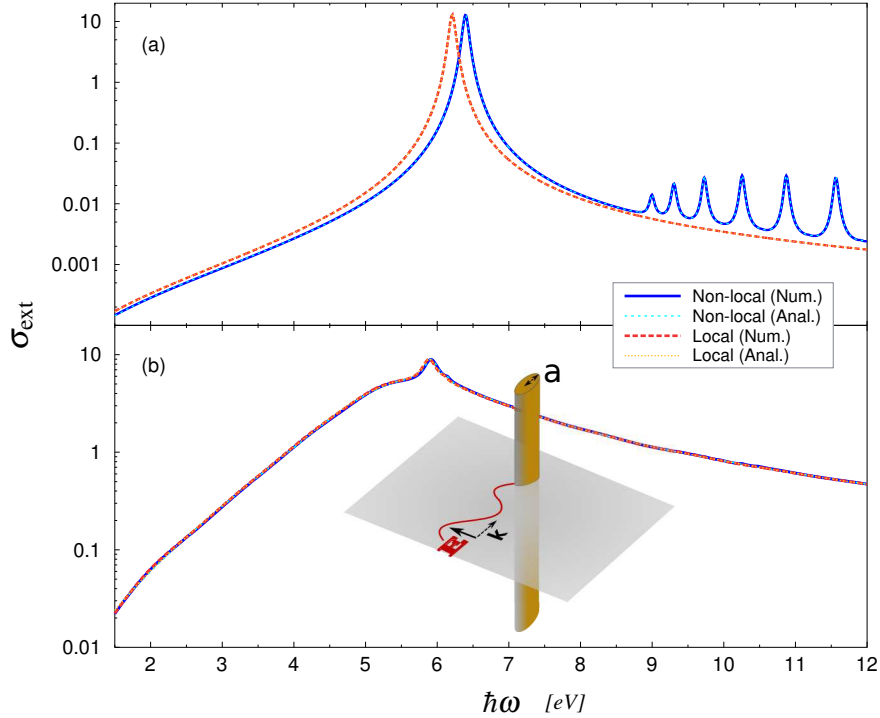}
\caption{Extinction-cross section $\sigma_{\text{ext}}$ (logarithmic scale) versus  frequency for cylindrical nanorods for two radii: (a) radius $a=2\,{\rm nm}$, (b) $a=25\,{\rm nm}$. $\sigma_{\text{ext}}$ is normalized to the diameter of the
rod.  Both panels show comparisons of
numerical simulations of Eq.~(\ref{eq:coupledequations}) to analytical results both for local response ($\beta = 0$) and for nonlocal response ($\beta = \sqrt{3/5} v_{\rm F}$). All numerical curves overlap the corresponding analytical curves.
\label{Fig1}}
\end{figure}

First we compare our numerical method against analytical results for a single cylindrical
nanowire, where the cylindrical symmetry allows for analytical solutions in terms of Bessel and Hankel
functions, both for local and for nonlocal response~\cite{Ruppin:2001a,Raza:2011a}.
Figure~\ref{Fig1}
summarizes the results of our benchmarking, and illustrates the strong dependence of the nonlocal effects on the subwavelength size of the nanostructure. The figure shows the dimensionless scaled
extinction cross section $\sigma_{\rm ext}$, defined as the cross section per length of the wire, divided by its diameter. In more detail, $\sigma_{\rm ext}= (|P_{\text{abs}}|+|P_{\text{scat}}|)/(2 a I_0)$, where $I_{0}$ represents the intensity of the incident plane wave, $P_{\text{abs}}$ the
absorbed and $P_{\text{scat}}$ the scattered power per length of the wire, and $2 a$ is its diameter. The powers are obtained by numerically integrating the Poynting vector on a circle surrounding the nanowire.

Figure~\ref{Fig1}(a) compares the numerical results for a nanowire of radius $a=2\,{\rm nm}$ to the exact analytical solution, Eq.~(4) in Ref.~\cite{Raza:2011a}, both for local and nonlocal response. For these tiny nanowires, nonlocal effect can be considerable. To give two examples, the relative difference of $\sigma_{\rm ext}$ for nonlocal against local response in the figure can be up to 15.33, and the resonant
frequency is  $6.20\,{\rm eV}$ for local and $6.39\,{\rm eV}$ for nonlocal response, a considerable blueshift of $\varDelta=0.19\,{\rm eV}$, being more than $2\,\%$ of the resonance frequency.

The analytical and numerical curves overlap almost completely for the local-response model, and likewise for the hydrodynamical nonlocal model. Hence only two of the four curves are visible. Thus our numerical model accurately captures the prominent effects of nonlocal response, namely the blueshift of the
surface-plasmon resonance $\omega_{\rm p}/\sqrt{2}$~\cite{Ruppin:2001a,Abajo:2008a,Raza:2011a,David:2011a}, as well as the confined bulk plasmon resonances above the plasma frequency $\omega_{\rm p}$~\cite{Ruppin:2001a,Raza:2011a}. More quantitatively, the relative error of the numerically computed cross section for nonlocal response is always smaller than $0.4\%$ in the entire frequency range of the figure, which includes many resonances, while for local response the relative error is always smaller than $0.6\%$. Further details about the accuracy and convergence of our numerical implementation are given in the Appendix.

Figure~\ref{Fig1}(b) shows the analogous four curves as Fig.~\ref{Fig1}(a), but now for a larger cylinder with radius $a=25\,{\rm nm}$. As one can see, the four curves all overlap almost completely. Over the frequency range 1.5 to $12\,{\rm eV}$, the maximum relative difference in cross section for the nonlocal against the local model is $7\%$, which means  that for most practical purposes nonlocal effects will be negligible for a single nanowire of this size. However, as we will see in Sec.~\ref{Sec:2cylinders}, this conclusion does not carry over to a dimer of two such nanowires! The maximum relative error in the numerically computed cross section for nonlocal response in Fig.~\ref{Fig1}(b) is $0.89\%$ while for local response it is $0.86\%$. These numbers show that the numerical method is accurate in a large frequency range also for larger nanostructures, here up to sizes where nonlocal effects can be neglected.

Summarizing, the implementation for the cylindrical nanowire is accurate in a wide range of frequencies and length scales, correctly reproducing the location of resonances and their amplitudes even for high frequencies beyond the plasma frequency.

\section{Nonlocal effects in dimers of cylindical nanowires}\label{Sec:2cylinders}
Having addressed single isolated nanowires, we now turn to nanowire dimers. For two closely separated
nanostructures, the usual local-response model predicts a strong hybridization of the surface-plasmon resonance~\cite{Prodan:2003a,Davis:2010a}. Thus dimers can even display resonances in the visible, with strongly
enhanced fields in the tiny gap separating the two parts. Here we study the effects of nonlocal response both on the hybridization and on the field enhancement. We consider dimers of cylindrical nanowires (in this section) and bow-tie nanowires in Sec.~\ref{Sec:bowtie}.

For definiteness, we consider a dimer of two identical cylindrical nanowires, with the same radius $a=25\,{\rm nm}$ as for the single nanowire of Fig.~\ref{Fig1}(b), and separated by a few-nm gap of size $d \ll a$. When excited by a plane wave normal to the wire axis and with the electrical field polarized across the dimer gap, the surface modes for this structure become highly confined to -- and enhanced in -- the narrow region of
the gap. This field configuration is due to the hybridization of the modes of the isolated
nanorods~\cite{Prodan:2003a,Brandl:2005a,Abajo:2008a}.

\begin{figure}[h!]
\centering \includegraphics{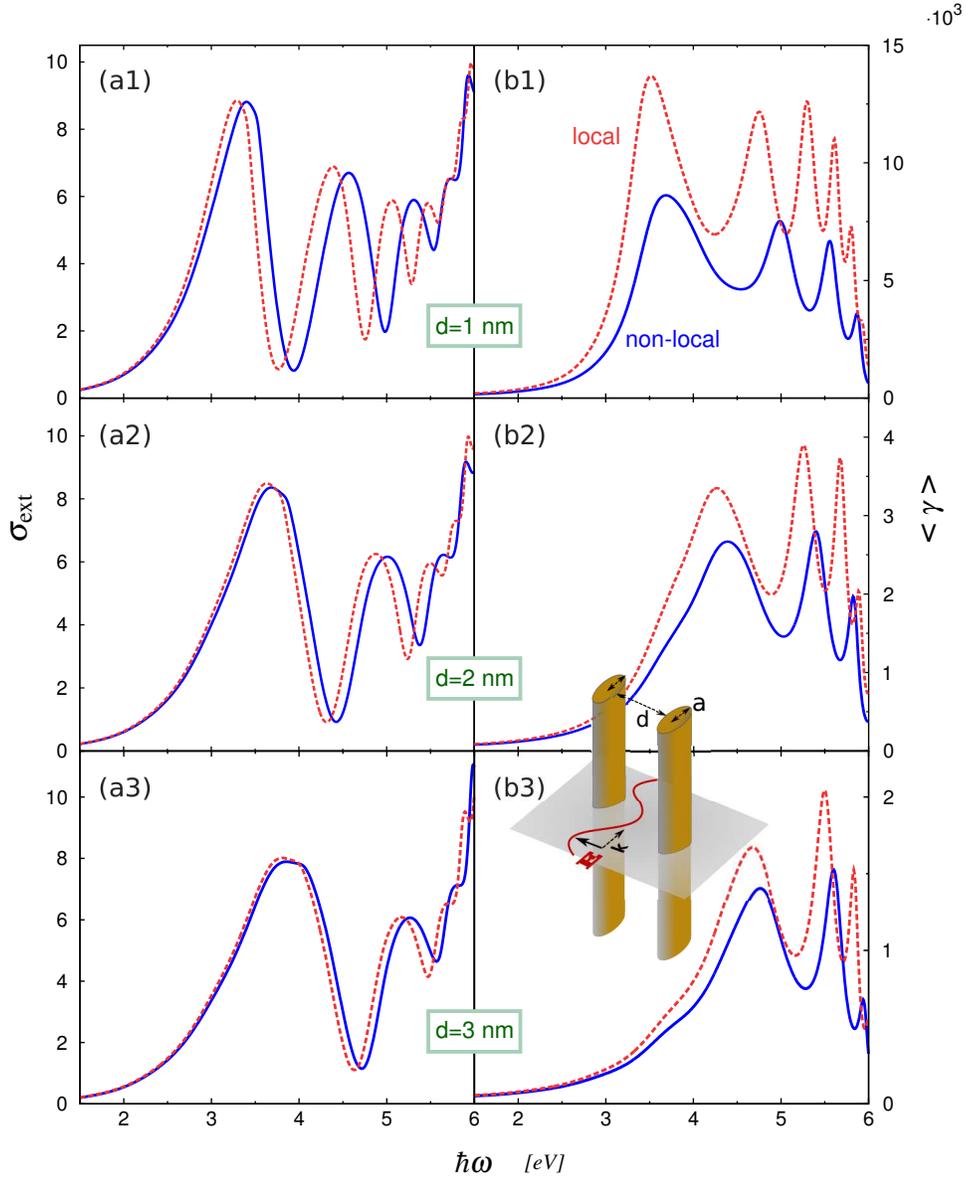}
\caption{Extinction-cross section versus frequency for a dimer of
two cylindrical nanorods of radius $a=25$\,nm separated by a distance $d$, excited by a TM-polarized plane wave with wave vector perpendicular to the line connecting the centers of the cylinders, as illustrated in the inset with $a/d$ not to scale. $\sigma_{\rm ext}$ is normalized to the diameter of the single wire. The left panels labeled (a) depict $\sigma_{\rm ext}$, while the  (b) panels on the right show the average field enhancement  $\langle\gamma\rangle$ as defined in the main text. 
The upper panels~(a1) and (b1) correspond to $d=1\,{\rm nm}$,
where the SPR appears at $3.28\,{\rm eV}$ and $3.40\,{\rm eV}$ for the local and nonlocal case, respectively.
The middle panels~(a2) and (b2)  concern $d=2$\,nm where the SPR appears at $3.63\,{\rm eV}$ in the local case and at $3.69\,{\rm eV}$ in the nonlocal
one. The lower panels~(a3) and (b3) correspond to $d=3\,{\rm nm}$. \label{Fig2}}
\end{figure}

We examine the effects of nonlocal response on energy confinement and field
enhancement for several gap sizes of 1 nm and beyond, where quantum tunneling, absent in our model, can be neglected~\cite{Zuloaga:2009a}.
The results for cylindrical dimers are summarized in Fig.~\ref{Fig2}.
We first discuss the three  panels~(a1-a3) on the left of Fig.~\ref{Fig2}, that depict extinction cross sections for increasing gaps, before addressing the average field enhancements in the right-hand panels.

The results in panel (a1) corresponding to a very small gap of $d=1\,{\rm nm}$ show a
strong plasmonic interaction between the two nanowires. This interaction gives rise to a pronounced hybridization and consequently the main surface-plasmon resonance (SPR) now appears at much lower energy of $3.28\,{\rm eV}$  for local response and $3.40\,{\rm eV}$ for nonlocal response, as
opposed to the SPR in an isolated wire around $5.91$\,eV in both local and nonlocal case, recall
Fig.~\ref{Fig1}(b). The important point is a pronounced nonlocal blueshift of $0.12\,{\rm eV}$ of the hybridized dimer resonance.
As the separation is doubled to $d=2\,{\rm nm}$, the hybridization decreases and consequently the SPR
appears at higher frequency around $3.63\,{\rm eV}$ for local and $3.69\,{\rm eV}$ for nonlocal response. Again the effect of nonlocal response is a noticeable blueshift of the hybridized dimer resonance, still noticeable but smaller than for $d=1\,{\rm nm}$.  Finally, for $d=3\,{\rm nm}$, the
hybridization is again weaker, so that the SPR is again blueshifted in the direction of the single-nanowire SPR (see Fig.~\ref{Fig1}). In all three cases, there is a nonlocal blueshift of the SP resonance frequency. This shift vanishes for larger $d$, because the blueshift for single nanowires with radius $a=25\,{\rm nm}$ also vanishes, as we saw in Fig.~\ref{Fig1}(b).

Now let us discuss the field enhancement in the right panels of Fig.~\ref{Fig2}, or more precisely the local field (intensity) enhancement factor $\gamma({\mathbf r})=|{\mathbf E}({\mathbf r})|^{2}/|E_{0}|^2$, where ${\mathbf E}({\mathbf r})$ is the local electrical field and $E_{0}$ the amplitude  of the incoming plane wave.
Dimers support modes that are strongly localized in the gap separating the two nanowires. Due to
the strong spatial localization, the amplitude of the local electrical field ${\mathbf E}({\mathbf r})$ may by far exceed the amplitude $E_{0}$ of the incoming plane wave.
Rather than considering local field enhancements in single points, we will consider spatially averages, because local field probes such as atoms cannot be positioned with infinite precision. The average field enhancements $\langle\gamma\rangle$ in Fig.~\ref{Fig2}(b) were obtained by line-averaging $\gamma({\bf r})$ over the narrow gap along the axis $\ell$ of the dimer, i.e.
\begin{equation}\label{gamma_av}
\langle\gamma\rangle = \frac{\int_\ell \mbox{d}{\mathbf r}\, \gamma({\mathbf r})}{\int_\ell \mbox{d}{\mathbf r}} =
\frac{1}{ E_0^2 d} \int_\ell \mbox{d}{\mathbf r}\, |{\mathbf E}({\mathbf r})|^2.
\end{equation}
 A direct comparison of left and right panels of Fig.~\ref{Fig2} shows that the spectral dependence of the field enhancement $\langle\gamma\rangle$ for lower frequencies is similar to the corresponding extinction cross section $\sigma_{\rm ext}$. However, $\langle\gamma\rangle$ peaks at higher frequencies than $\sigma_{\rm ext}$, both for local and nonlocal response. The agreement between the two types of curves is not complete, because the extinction cross section is the sum of a scattering and an absorption cross section. The latter can be interpreted as a two-dimensional loss average {\em inside} the cylinders, whereas the field enhancement factor $\langle\gamma\rangle$ of Eq.~(\ref{gamma_av})  is a more local one-dimensional spatial average of the empty space  {\em in between} the cylinders.
 For that reason,  the extinction $\sigma_{\rm ext}$ can be high near $6\,{\rm eV}$ while the field enhancement $\langle \gamma \rangle$ is low. Indeed, there exists a resonant hybridized mode at this frequency (hence the peak in $\sigma_{\rm ext}$) with a mode profile with low intensity on the line joining the cylinder centers (which explains the low value for $\langle \gamma \rangle$); the mode intensity grows away from this line (mode profile not shown).

For the narrowest gap of $d=1\,{\rm nm}$, Fig.~\ref{Fig2}(b1) shows a field enhancement of $\langle\gamma\rangle= 1.4\times10^4$ at the main SPR (i.e. at the lowest-energy peaks) for local response, and
of $\langle\gamma\rangle=
8.6\times10^3$ for nonlocal response. Thus, the nonlocal response gives a strong average field enhancement, yet it is considerably smaller than for local response. As the gap distance increases to $d=2$\,nm, the maxima of $\langle \gamma\rangle$  decrease for both types of material response.
Finally  in Fig.\ref{Fig2}(b3) the gap has been increased further to $d=3\,{\rm nm}$, and we still find considerable differences for nonlocal response, even though there is a further decrease both of the peak amplitudes  of $\langle \gamma\rangle$
and of their relative differences.

Thus we find for the scattering off cylindrical nanowire dimers with radius  $a=25\,{\rm nm}$ separated by 1 to 3~nm  that nonlocal response effects are considerable, for cross sections but more so for field enhancements, even though in Sec.~\ref{Sec:1cylinder} we found that nonlocal effects were negligible for the single cylindrical nanowire with the same radius.

The above discussion compares nonlocal with local peak enhancements reduced amplitudes. Since the nonlocal peaks also shift in frequency, the nonlocal effects are even  more pronounced when studying observables  at fixed frequencies. For then Fig.~\ref{Fig2}(a1) shows that $\sigma_{\rm ext}$  can be reduced by a factor of 3.1 at 4.0 eV, and enhanced by a factor of 3.4 at 3.7 eV. Nonlocal field enhancements for some frequencies can also turn out to be just a bit larger, for example by a factor 1.34 at 5.95 eV in Fig.~\ref{Fig2}(b3), even though we mostly find smaller nonlocal values for the field enhancement $\langle \gamma\rangle$ in between the cylinders. Finally, it is interesting to notice frequencies in Fig.~\ref{Fig2}(b) at which the local-response field enhancement peaks while the nonlocal-response field enhancement goes through a minimum.

\section{Nonlocal effects in bow-tie nanowires}\label{Sec:bowtie}
\begin{figure}[h!]
\centering \includegraphics{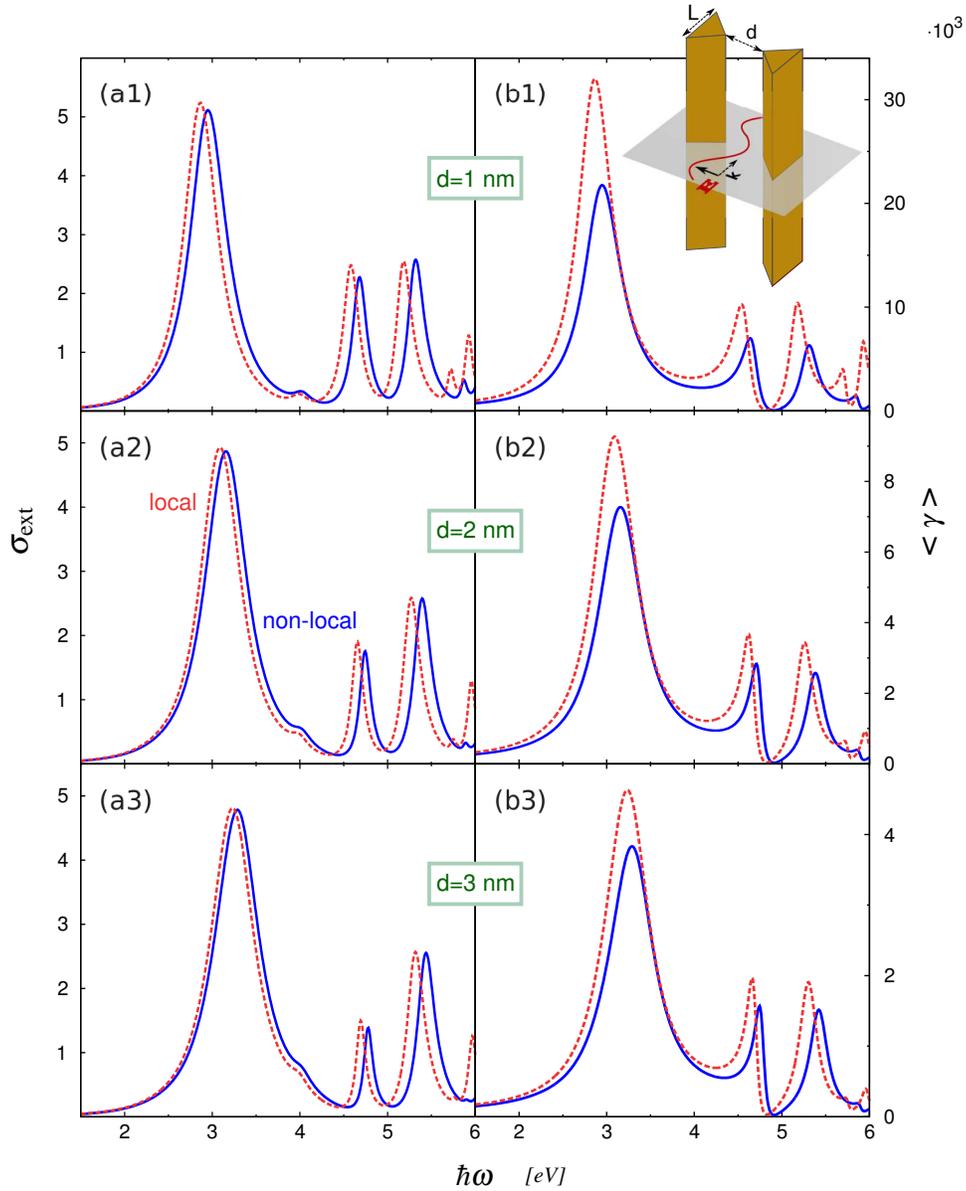}
\caption{Extinction-cross section versus frequency for a dimer of
two equilateral triangles nanowires of side $L=45$\,nm separated by a distance $d$, excited by perpendicularly incident TM-polarized light. The inset shows a sketch with $d/L$ not to scale. As in Fig.~\ref{Fig2}, the left and right panels show extinction cross sections $\sigma_{\rm ext}$ and average field enhancements $\langle \gamma\rangle$, respectively. Upper, middle, and lower panels again correspond to $d=1,2$ and $3\,{\rm nm}$.  \label{Fig3}}
\end{figure}

Let us now consider light scattering off  the bow-tie nanowire dimer, the geometry as sketched in the inset of Fig.~\ref{Fig3}.
Bow-tie structures can give rise to high field enhancement near the almost touching sharp tips of the two triangles~\cite{Merlein:2008a,Marty:2010a,Ding:2010a,Barnard:2011a}. Sharp features in nanoplasmonic structures are known to give strong field enhancements. The cylindrical dimer of the previous section did not have this type of field enhancement, and therefore it is interesting to compare nonlocal effects on field enhancement for the two types of dimers.

In particular, we consider a dimer of two equilateral triangular nanowires with side $L=45\,{\rm nm}$.
The tips of the triangles have been rounded  with a radius of curvature of $1$\,nm. Such rounding for computational reasons is common practice~\cite{Giannini:2007a,Wallen:2008a,Cui:2010a}.
In Fig.~\ref{Fig3} we present extinction cross sections and average field enhancements for bow-tie dimers, analogous to Fig.~\ref{Fig2}. The field enhancements are averaged over the line connecting the almost touching tips of the triangles.

The results resemble those for the circular dimers, but with narrower and thus less overlapping plasmon resonances. One main feature is again that all (hybridized) surface-plasmon resonances are blueshifted due to nonlocal response.  The case $d=1$\,nm shows a strong
interaction between the plasmons localized on the tip of the triangles, and the main SPR appears at
$2.86$\,eV for local and $2.96$\,eV for nonlocal response. Larger distances imply weaker hybridization so that lowest-energy resonances  shift to higher energies. Indeed for $d=2\,{\rm nm}$ the resonance frequency of the main SPR is $3.08$\,eV in the
local model but blueshifted by  $0.07$\,eV in the nonlocal one. Finally for $d=3\,{\rm nm}$ the main SPR occurs at $3.24$\,eV in the
local case and $3.29$\,eV in the nonlocal one. As for the dimer of cylinders, for the bow-tie nanowire dimer we find larger nonlocal blueshifts of peaks in $\sigma_{\rm ext}$ in case of larger hybridization of surface-plasmon resonances.

As to the importance of nonlocal effects in the field enhancement for the bow-tie dimer, Fig.~\ref{Fig3}(b) shows that $\langle\gamma\rangle$ peaks at $3.2\times10^4$ in the local case and at $2.17\times10^4$ in the
nonlocal case for $d=1$\,nm. For the tip distance $d=2$\,nm these values are $9.2\times10^3$ and $ 7.2\times10^3$, respectively, and they become $4.7\times10^3$ and $4.6\times10^3$  for $d=3\,{\rm nm}$. For field enhancement, like for cross sections, nonlocal effects turn out to be more important in case of larger hybridization.

If instead of focusing on peak values we compare again nonlocal with local response at fixed frequencies, then for bow-tie dimers we can easily identify frequency intervals for which the field enhancement is larger for nonlocal than for local response, in contrast to what we found for the cylindrical dimers. For $d=3$ nm at 4.81 eV, $\langle\gamma\rangle$ is even 14.3 times larger for nonlocal than for local response.    And for some frequencies, the nonlocal cross section is larger while for other frequencies it is smaller than the local response, and all differences are roughly within a factor of 5. Near 4.8 eV the field enhancements vanish even though the cross section peaks. Again this combination is a fingerprint of  a resonant mode with low mode density near the center of the dimer.

The peak field enhancements $\langle \gamma\rangle$ for the bow-tie dimers in Fig.~\ref{Fig3}(b) are roughly a factor 2 higher than for the cylindrical dimers in Fig.~\ref{Fig2}(b), both for local and for nonlocal response. Thus the quite different shapes of the dimers give rise to non-negligible but not too big differences in field enhancement, considering that all enhancement peaks  $\langle \gamma\rangle$ are of order $10^{4}$ or higher. This applies both to local and to nonlocal response.

\section{Discussion and conclusions}\label{Sec:Conclusions}
We have implemented the hydrodynamical Drude model for arbitrary nanowire geometries as an extension of state-of-the-art numerical software in nanoplasmonics. Our code was tested against analytical results and was shown to be very accurate. We advocate the use of such benchmark problems, to be able to present  results for more complex geometries with confidence.

We studied tiny cylindrical nanowires, their dimers, as well as bow-tie dimers. In all cases we find that (hybridized) surface-plasmon resonances are blue-shifted due to nonlocal response. It is not simply the size of the plasmonic nanostructure that determines whether nonlocal effects are important. For example, we found that nonlocal effects were negligible for the extinction cross section of $25\,{\rm nm}$ cylinders, but important for their closely spaced dimers.

We find that the usual explanations of plasmonic hybridization carry over to nonlocal theories. Below the plasma frequency we do not find new resonances due to nonlocal response, in agreement with our Ref.~\cite{Raza:2011a},  but the hybridized resonances occur at higher frequencies than expected based on a local-response picture, with modified mode profiles. The nonlocal blueshifts are a correction to the larger hybridization energies. As an important conclusion, we find that  the {\em nonlocal blueshifts are larger for more strongly hybridized plasmonic structures}. It is a package deal, so to say: one cannot have the one without the other, at least for dimers for which the individual parts are too big to exhibit any nonlocal blueshift.

Bow-tie dimers have sharp tips and cylindrical dimers do not, but this did not result in large differences in nonlocal shifts in both cases. For the extinction cross sections of the dimers we found blueshifts in resonance peaks but hardly a change in their amplitudes.  For the average field enhancements on the other hand, we find both blueshifts and a reduction in height of resonance peaks, roughly by a factor of 2 for cylindrical dimers and a factor of 1.5 for the bow-tie nanowires. Nevertheless this general conclusion is fully consistent with the fact that for some fixed frequencies, the field intensities in between the cylinders or bow-ties are extra enhanced due to nonlocal as compared to local response, by up to factors of 14. We also found frequencies for which the field enhancement peaks for nonlocal response but has  minimum for local response. Thus it is important to take effects of nonlocal response into account in the context of spontaneous emission rates of nearby quantum dot emitters, fluorescence of dye molecules, and
surface-enhanced Raman scattering (SERS) of bio-molecules, especially  in the close vicinity of strongly hybridized plasmonic nanostructures.

In this work we neglected quantum tunneling effects~\cite{Zuloaga:2009a,Oeztuerk:2011a}. In their quantum many-body calculations, Zuloaga {\em et al.}~\cite{Zuloaga:2009a} identify a cross-over regime for dimer gaps between 0.5 and $1.0\,{\rm nm}$, where narrow-barrier quantum tunneling effects strongly reduce the classical hybridization energies, and a conductive regime for $d<0.5\,{\rm nm}$. The classical (local-response) limit is also found in their calculations for large dimer separations. The message of our hydrodynamical calculations is that significant departures from classical local-response theory will already occur at larger dimer separations  in the range  $1$ to $10\,{\rm nm}$, where quantum tunneling between the dimers is negligible, and that the local-response limit is found for large separations (i.e. for the individual $a=25\,{\rm nm}$ cylindrical wires for example). It would be gratifying to see experimental nonlocal blueshifts for dimers, and to see many-body quantum calculations confirming not only the large-separation local-response limit, but also the hydrodynamical nonlocal-response blueshifts for dimers with separations in the range 1-3 nm as presented here.

\section{Acknowledgments}
We thank Sanshui Xiao for useful discussions.
This work was financially supported by the Danish Research Council for Technology and Production
Sciences (Grant No. 274-07-0080), and by the FiDiPro program of the Finnish Academy.

\section{Appendix: hardware implementation and accuracy studies}
All the presented results are obtained on a personal computer equipped
with four Intel 2Ghz processors and 24Gb of RAM. Single-frequency calculations typically take about one minute for regular geometries with smooth boundaries. The frequency step used to make the spectra is $0.0088\,{\rm eV}$, and with this value all the data presented can be harvested in less than three days on our pc.

\begin{figure}[h!]
\centering
\includegraphics[scale=0.9]{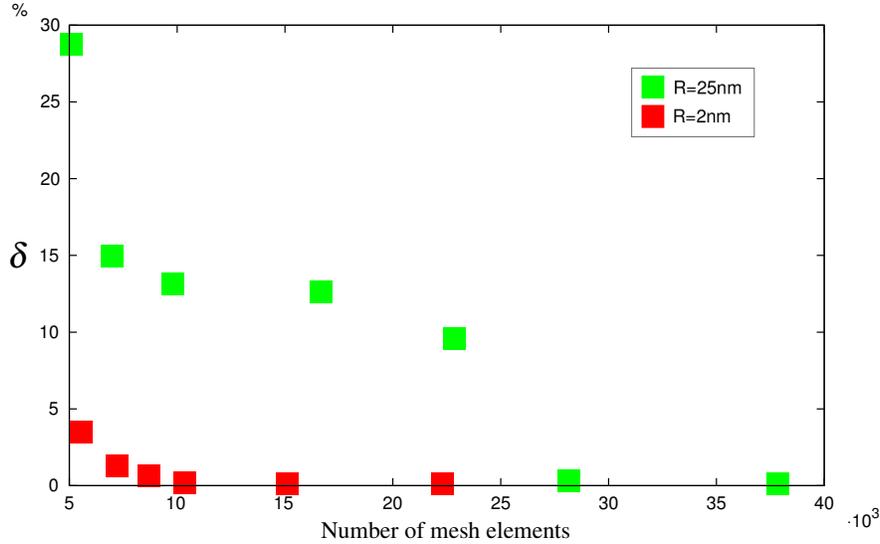}
\caption{Relative error $\delta$ of the numerically calculated extinction cross section $\sigma_{\rm ext}^{\rm num.}$ for a single cylinder of radius $a=2\,{\rm nm}$ at the nonlocal-response resonant frequency $\omega_{\text{sp}}=6.38$\,eV, and for a single cylinder of radius $a=25\,{\rm nm}$ at the frequency $5.91$\,eV, as a function of the number of mesh elements. Modeled with parameters of Au as given in Sec.~\ref{Sec:theory}.
\label{Fig4}}
\end{figure}

The code that we have developed has undergone several reliability tests. In Fig.~\ref{Fig4} we present a convergence study that has been carried out for the nanorod of
radius $a=2$\,nm with nonlocal response. The simulation box is a square of side $W=300$\,nm, surrounded by a PML that is $30$\,nm thick. The walls of the box are in the far-field zone of the scatterer, so that they cannot influence the field scattered away from our system under test.

First we studied the convergence of the nonlocally blueshifted value of the surface-plasmon resonance frequency, and we found that it converged to the analytical value $6.38$\,eV without much mesh refinement. Then at this resonance frequency, being a challenging spot,  we studied the dimensionless convergence parameter $\delta =|\sigma_{\text{ext}}^{\text{num}}-\sigma_{\text{ext}}^{\text{an}}|/\sigma_{\text{ext}}^{\text
{an}}$
as a function of the number of mesh elements used in the calculation. Meanwhile the number of mesh elements of both the box and the PML were kept fixed. The parameter $\delta$ represents the
relative error of $\sigma_{\text{ext}}^{\text{num}}$ with respect to the
$\sigma_{\text{ext}}^{\text{an}}$, so that a vanishing $\delta$ not just signifies convergence but rather convergence to the analytical value.

Fig.~\ref{Fig4} depicts $\delta$ versus the number of mesh elements. It shows that the convergence sets in for less than 8000 mesh elements,  a small number for modern pc's. In particular, $\delta= 1.3\,\%$ for 7241 mesh elements and $0.12\,\%$ for 15106 mesh elements. This means that our code could
easily run on a laptop, at least for the small structures considered here. The non-uniform mesh size allows also convergence on a pc for a larger cylinder with $a=25\,{\rm nm}$, with 	$\delta=0.11\%$ for  37830 mesh elements. Much larger structures can of course be handled on a powerful workstation.

\end{document}